\documentclass[aps,prb,twocolumn,groupedaddress]{revtex4}%
\usepackage{amsmath}
\usepackage{graphics}
\usepackage{graphicx}
\usepackage{epsfig}
\usepackage{amssymb}
\usepackage{amsmath}
\usepackage{amsfonts}%
\providecommand{\U}[1]{\protect\rule{.1in}{.1in}}

\begin{document}
\title{Superconductivity in strongly correlated systems for local repulsive interactions}
\author{Humberto M. \surname{Silva}$^{1}$}
\affiliation{$^{1}$Universidad Nacional de Ingenier\'{i}a, Av. Tupac Amar\'{u} 210, Rímac 15333, Lima, Per\'{u}.}
\author{Francisco \surname{Din\'{o}la Neto}$^{2}$}
\email{dinola@ufam.edu.br}
\author{Griffith M. A. R.$^{2}$}
\author{Minos A. Neto$^{2}$}
\author{Octavio D. R. Salmon$^{2}$}
\affiliation{$^{2}$Universidade Federal do Amazonas - UFAM, Manaus
69077-000, AM, Brazil}
\author{Mucio A. Continentino$^{3}$}
\author{Amos Troper$^{3}$}
\affiliation{$^{3}$ Centro Brasileiro de Pesquisas F\'isicas, Rio de Janeiro 22290-180, RJ, Brazil}

\begin{abstract}The understanding of the mechanisms responsible for superconductivity in strongly correlated
systems is an interesting and important subject in condensed matter physics. Several theoretical
proposals were considered for these systems. The Coulomb interaction between electrons allow a
new approach to study this problem. In this paper, we use a usual Hubbard model with a local
repulsive interaction to describe a 2D system. The system of equations are solved using the Green’s
functions method, within a Hubbard-I mean field approximation, which allows to treat the strong
interaction limit. We consider both cases of attractive and repulsive interactions and obtain the
zero temperature phase diagram of the model. Our results show, in the repulsive case, the existence
of a superconducting ground state mediated by the kinetic electronic energy and described by a
non-local order parameter. A minimum value of the repulsive interaction $U_{min}$ is required to create
a pairing state. At finite temperatures, for strong interactions, the critical temperature $T_c$ shows a
saturation similar to the Bose-Einstein condensation observed for strong attractive interactions.
\end{abstract}
\maketitle

\section{Introduction}

The BCS theory\cite{bcs} is very successful to explain the
essence of the superconductivity phenomena in superconducting
metals and alloys. However, since the discovery
of superconductivity in the cuprate compounds and other
High $T_c$ materials\cite{HTSC}, experimental evidences suggest that
they do not obey the BCS mechanism of pairing. This
fact indicates the possibility of a non-phononic pairing in
these systems, but establishes that the BCS key mechanism
of electron pairing is still valid\cite{Gough}. Anderson\cite{Anderson} was
the first to propose that the essence of high-temperature
superconductivity is contained in the 2D Hubbard model
for a square lattice with a repulsive interaction $U$.

One attempt to study the High $T_c$ superconductors in
copper oxides is from the point of view of the strength
of the interactions forming the Cooper pairs. First of all
it is necessary to understand the so-called crossover between
the BCS limit of the weak coupling regime and the
BEC (Bose-Einstein condensation) limit of strong coupling
interactions. The first limit describes a superfluid
composed of fermions with arbitrarily weak attraction
and the latter a condensed system of diatomic molecules.
Nozi\`{e}res and Schimtt-Rink\cite{nozieres} in 1985, were the first to
study this problem by introducing a finite temperature
in the theoretical analysis. Using a diagrammatic formulation
they showed that, the transition temperature
Tc (temperature at which the material becomes superconducting)
evolves smoothly as a function of coupling.
Later, S\'{a} de Melo \emph{et al.}\cite{samelo} in 1993, reshaped the work of
Nozi\`{e}res and raised the possibility that the physics of the
crossover BCS-BEC may be relevant to the mechanism
of high temperature superconductors.

In this work we propose to study the superconductivity in the repulsive Hubbard model for a 2D system using the Green’s functions method similarly as made by Beenen\cite{Beenen} and Sarasua\cite{Sarasua}. Recently, Lisandrini \emph{et al.}\cite{Lisandrini} studied superconductivity in the repulsive Hubbard model induced by density-assisted hopping. Generally the Hubbard model is used to treat the superconductivity for the attractive case. The repulsive term is often used to justify some experimental observations in the strongly correlated electronic systems (SCES). However the repulsive character of the interactions is not associated with the appearance of the superconducting state\cite{bastide,dinola}. In our approach, we treat the second order propagators introducing the Hubbard-I approximation, similarly as made by Caixeiro and Troper\cite{caixeiro}. The Hubbard Hamiltonian studied in the present work contains just a local interaction between fermions differently from Ref\cite{caixeiro} where non-local interaction were considered. Nevertheless the Hubbard-I approximation gives rise to a non-local superconducting gap amplitude related to the dispersion of the quasi-particles in the system besides a local gap term. We conclude that depending on the correlations that we consider the behavior of the system is controlled by the induced non-local characteristics.

As we have pointed out, we consider a Green’s functions method in order to obtain both the gap (local and non-local) and number equations. These equations are necessary to study self-consistently the behavior of the
superconducting gap and chemical potential as functions of the interaction strength. We discuss about the appearance of a critical minimum interaction $U_{min}$ at zero temperature ($T = 0$), required to give rise to a pairing state. For strong repulsive interactions we consider that the local gap amplitude is almost null remaining only the non-local term. In the opposite scenario for weak interactions we consider that only the local pairing remains. We can study also the finite temperature effects and the behavior of the critical temperature $T_c$. We consider here the case of a 2D system. For strong repulsive correlation it is observed an interesting saturation of $T_c$ similar to that observed in the BEC regime for attractive interactions.

The paper is organized as follows: In Sec. II we introduce the model and general formalism leading to the main equations within the Hubbard-I approximation. The goal is to analyze the energy of the quasi-particle excitations and obtain the self-consistent gap and total number equations. In Sec III we will study the strong coupling regime for the equations found in section II. In section IV we discuss the strong coupling limit of attractive
interactions where the non local gap amplitude vanishes. Finally, in section V conclusions and final comments are made.

\section{The Hubbard model}

We study the dynamics of the system through the Hubbard Hamiltonian
with local interaction. It is given by:

\begin{align} \label{1} \mathcal{{H}}
=\!\underset{ij\sigma}{\sum}t_{ij}^{d}
d_{i\sigma}^{\dag}d_{j\sigma}+\frac{U}{2}\underset{i\sigma}{\sum}n_{i\sigma}^{d}n_{i-\sigma}^{d}-\mu\underset{i\sigma}{\sum}n_{i\sigma}^{d}
\end{align}
where $d_{i\sigma}^{\dag}$($d_{j\sigma}$) are the fermionic creation
(annihilation) operators at site $i$ for $d$-electrons respectively,
with spin $\sigma$ up($\uparrow$) or down($\downarrow$). The density
operator is given by $n_{i\sigma}^d=d_{i\sigma}^{\dag}d_{i\sigma}$,
$t_{ij}^{d}$ is the hopping integral between $i$ and $j$
nearest-neighbors sites for $d$ electrons, $U$ is the effective
potential between $d$ electrons and $\mu$ is the chemical potential.
We propose in the Hamiltonian in Eq.(\ref{1}) to describe the
formation of $d$-$d$ Cooper pairs via self-consistent equations for
the density of electrons and gap amplitude. The critical temperature
$T_{c}$ and the superconducting phase diagrams as function of
the parameters of the model are obtained straightforward from these
equations.

The superconducting order parameter and number
equation are obtained from the normal $\langle\langle
d_{j\sigma};d_{i\sigma}^{\dag}\rangle\rangle$ and anomalous $\langle\langle
d_{j-\sigma}^{\dag};d_{i\sigma}^{\dag}\rangle\rangle$ Green’s functions obtained using the equations of motion method, within the Hubbard-
I approximation. We calculate the equations
of motion for the highest order new generated propagators
$\langle\langle
n_{j\sigma}^{d}d_{j-\sigma}^{\dag};d_{i\sigma}^{\dag}\rangle\rangle$
and $\langle\langle
n_{j-\sigma}^{d}d_{j\sigma};d_{i\sigma}^{\dag}\rangle\rangle$ that
results in

\begin{align}\label{2}
(\omega\!-\!U)\langle\langle
n_{j-\sigma}^{d}d_{j\sigma};d_{i\sigma}^{\dag}\rangle\rangle\!=\nonumber \\
\frac{\langle
n_{i\!-\!\sigma}^{d}\rangle}{2\pi}\delta_{ij}\!+\!\underset{l}{\sum}({t}_{lj}^{d}-\mu\delta_{lj})\langle\langle
n_{j-\sigma}^{d}d_{l\sigma};d_{i\sigma}^{\dag}\rangle\rangle
\end{align}
and
\begin{align}\label{3}
(\omega\!+\!U)\langle\langle
n_{j\sigma}^{d}d_{j-\sigma}^{\dag};d_{i\sigma}^{\dag}\rangle\rangle\!=\nonumber
\\ \!\frac{\langle
d_{j\sigma}^{\dag}d_{j-\sigma}^{\dag}\rangle}{2\pi}\delta_{ij}\!-\!\underset{l}{\sum}({t}_{lj}^{d}-\mu\delta_{lj})\langle\langle
n_{j\sigma}^{d}d_{l-\sigma}^{\dag};d_{i\sigma}^{\dag}\rangle\rangle,
\end{align}
where was previously performed the Hubbard-I
approximation\cite{hubbard}

\begin{align}\label{4}
\underset{l}\sum
({t}_{lj}^{d}-\mu\delta_{lj})\langle\langle\left(d_{j\sigma}^{\dag}d_{l\sigma}-d_{l\sigma}^{\dag}
d_{j\sigma}\right)d_{j-\sigma}^{\dag};d_{i\sigma}^{\dag}\rangle\rangle\simeq0,
\end{align}
and

\begin{align}\label{5}
\underset{l}\sum
({t}_{lj}^{d}-\mu\delta_{lj})\langle\langle\left(d_{j-\sigma}^{\dag}d_{l-\sigma}-d_{l-\sigma}^{\dag}
d_{j-\sigma}\right)d_{j\sigma};d_{i\sigma}^{\dag}\rangle\rangle\simeq0.
\end{align}

To decouple the highest order propagators and obtain a closed system
of equations we consider that\cite{japiassu}
\begin{align}\label{6}
\langle\langle
n_{j-\sigma}^{d}d_{l\sigma};d_{i\sigma}^{\dag}\rangle\rangle =
\langle n_{-\sigma}^{d}\rangle\langle\langle
d_{l\sigma};d_{i\sigma}^{\dag}\rangle\rangle\nonumber\\ + \langle
d_{j-\sigma}d_{l\sigma}\rangle\langle\langle
d_{j-\sigma}^{\dag};d_{i\sigma}^{\dag}\rangle\rangle
\end{align}
and
\begin{align}\label{7}
\langle\langle
n_{j\sigma}^{d}d_{l-\sigma}^{\dag};d_{i\sigma}^{\dag}\rangle\rangle
= \langle n_{\sigma}^{d}\rangle\langle\langle
d_{l-\sigma}^{\dag};d_{i\sigma}^{\dag}\rangle\rangle\nonumber\\ -
\langle d_{j\sigma}^{\dag}d_{l-\sigma}^{\dag}\rangle\langle\langle
d_{j\sigma};d_{i\sigma}^{\dag}\rangle\rangle.
\end{align}

Next step is to perform the Fourier transformation. At this point, we
must be careful about the transformation in terms like,

\begin{align}\label{8}
U\underset{ijl}\sum({t}_{lj}^{d}-\mu\delta_{lj})\langle
d_{l\sigma}^{\dag}d_{j-\sigma}^{\dag}\rangle\langle\langle
d_{j\sigma};d_{i\sigma}^{\dag}\rangle\rangle\,
\end{align}
that yelds
\begin{align}\label{9}
\left(U\Delta_{nl}-\mu\Delta\right)\langle\langle
d_{k\sigma};d_{k\sigma}^{\dag}\rangle\rangle.
\end{align}
The SC gap functions are given by
\begin{align}\label{10}
\Delta=U\underset{k}\sum\langle
d_{k\sigma}^{\dag}d_{-k-\sigma}^{\dag}\rangle
\end{align}
and the non-local gap amplitude ($\Delta_{nl}$)
\begin{align}\label{11}
\Delta_{nl}=\underset{k}\sum\epsilon_{k}\langle
d_{k\sigma}^{\dag}d_{-k-\sigma}^{\dag}\rangle
\end{align}
where $\epsilon_{k}=\underset{j}\sum
t_{lj}^{d}\textbf{e}^{ik(r_{j}-r_{l})}$.

Notice that the gap function $\Delta$ is related to local pairing
$\langle d_{i\sigma}^{\dag}d_{i-\sigma}^{\dag}\rangle$, while $\Delta_{nl}$ represents a non-local ($nl$) $\langle
d_{i\sigma}^{\dag}d_{j-\sigma}^{\dag}\rangle$ pairing amplitude.

In what follows we calculate the propagators and get

\begin{align}\label{12}
\langle\langle
d_{k\sigma};d_{k\sigma}^{\dag}\rangle\rangle_{\omega}\!=\!\frac{1}{2\pi}\left[\frac{\omega(\omega+U)(\omega-\tilde{U})\!+\!
\xi_{k}(\omega^2\!-\!\tilde{U}^2)}{P(\omega)}\right.\nonumber\\-\left.\frac{\!U\Delta\Delta_{nl}\!-\mu\Delta^2}{P(\omega)}\right]
\end{align}
and
\begin{align}\label{13}
\langle\langle
d_{-k-\sigma}^{\dag};d_{k\sigma}^{\dag}\rangle\rangle_{\omega}=
-\frac{1}{2\pi}\left\{\frac{\Delta[\omega(\omega-U)-\xi_{k}(\omega-\tilde{U})]}{P(\omega)}\right.\nonumber\\
+\left.\frac{(\omega-\tilde{U})(U\Delta_{nl}-\mu\Delta)}{P(\omega)}\right\}
\end{align}
where $\tilde{U}=U(1-n_{d}/2)$.

In our description we restrict the solutions to the non-magnetic case
i.e. $\langle n_{\sigma}^{d}\rangle=\langle
n_{-\sigma}^{d}\rangle=n_{d}/2$. The dispersion relation $\xi_{k}=\epsilon_{k}-\mu$, with
$\epsilon_{k}=k^{2}/2m$, for an electron gas with electronic mass $m$. Finally,
$P(\omega)$ is the $4^{th}$ order polynomial
\begin{align}\label{14}
P(\omega)=\omega^4-(\xi_{k}^2+U^2+\xi_{k}Un_{d})\omega^2
\nonumber \\
+[(U\xi_{k})^2\left(1-n_{d}/2\right)^2+(U\Delta_{nl}-\mu\Delta)^2].
\end{align}
The roots of $P(\omega)$ yield the poles of the Green's functions and
determine the energy of the excitations of the system. They are given by,
\begin{align}\label{15}
\omega_{(1,2)k}=\sqrt{A(k)\pm\sqrt{B(k)}},
\end{align}
where
\begin{align}
A(k)=\frac{\xi_{k}^2+U^2+\xi_{k}Un_{d}}{2}\nonumber
\end{align}
and
\begin{align}
B(k)=A(k)^{2}-\left[(U\xi_{k})^2\left(1-\frac{n_{d}}{2}\right)^2+(U\Delta_{nl}-\mu\Delta)^2\right].\nonumber
\end{align}

\begin{figure}[th] \centering
\includegraphics[angle=0,scale=1.0,height=7.5cm]{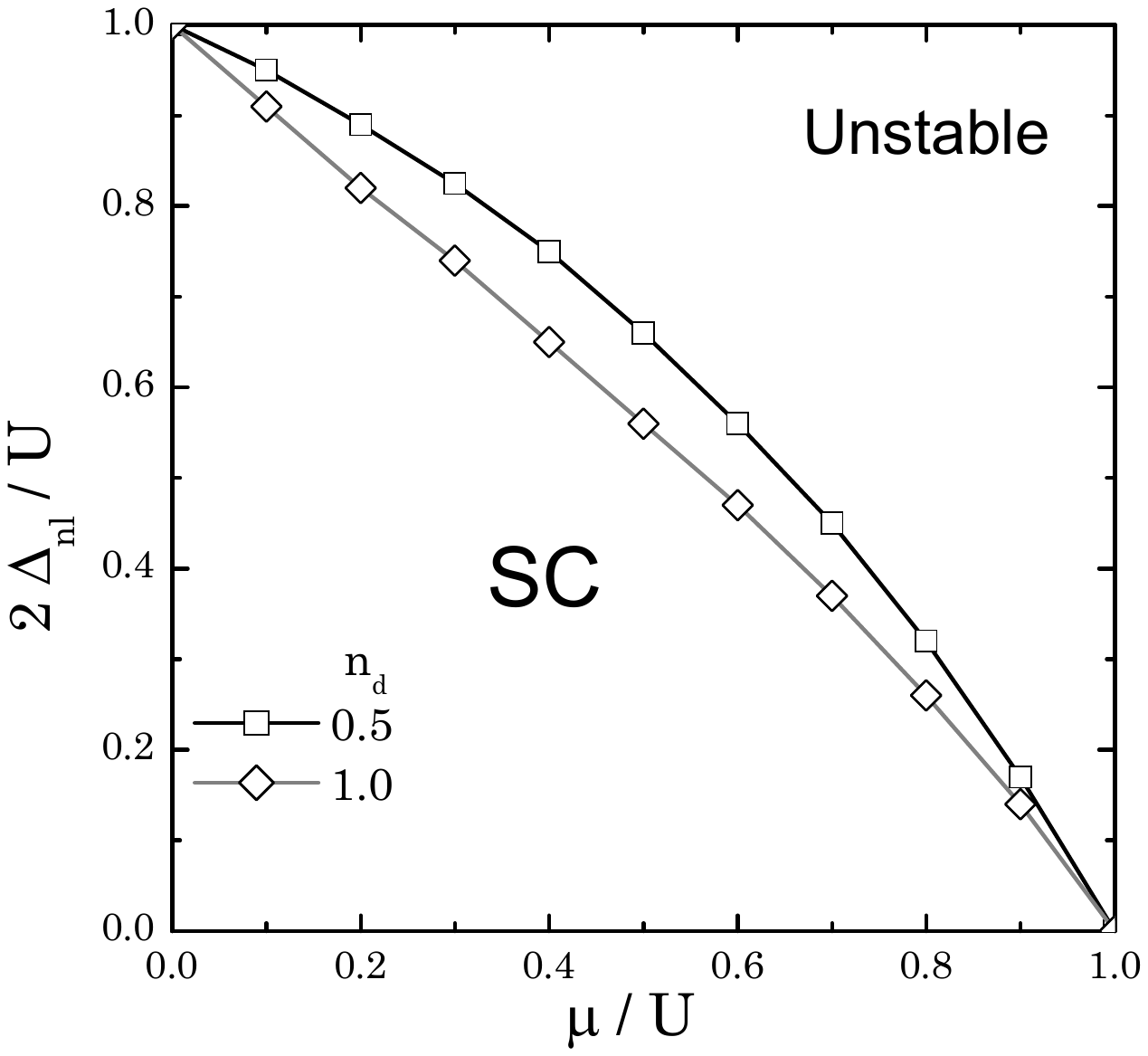}
\caption{Values of the parameters for which the dispersion relations of the quasi-particles are real, for $n_d = 0.5$ upper curve and $n_d = 1$. For values of $\Delta_{nl}$ above the lines superconductivity
is unstable as evidenced by an imaginary part in
the energy of the quasi-particle excitations.}\label{fig1}
\end{figure}

Next, with the propagators in Eqs. (\ref{12}) and (\ref{13}) and using the fluctuation-dissipation theorem, we calculate
the correlation functions and obtain a system of three equations from
Eqs. (\ref{10}), (\ref{11}),

\begin{align}\label{16}
n_{d}\!=\!\frac{1}{2}\underset{k}\sum\left[1\!-\!\frac{1}{(\omega_{1k}^2\!-\!\omega_{2k}^2)}
\underset{i=1}{\overset{2}{\sum}}(-1)^{i-1}\mathfrak{F}_{i}(\Delta,\Delta_{nl},n_{d})\right],
\end{align}
\begin{align}\label{17}
\Delta=U\underset{k}\sum\frac{1}{2(\omega_{1k}^2\!-\!\omega_{2k}^2)}
\underset{i=1}{\overset{2}{\sum}}(-1)^{i-1}
\mathfrak{G}_{i}(\Delta,\Delta_{nl},n_{d}),
\end{align}
and
\begin{align}\label{18}
\Delta_{nl}=\underset{k}\sum\frac{\epsilon_{k}}{2(\omega_{1k}^2\!-\!\omega_{2k}^2)}
\underset{i=1}{\overset{2}{\sum}}(-1)^{i-1}
\mathfrak{G}_{i}(\Delta,\Delta_{nl},n_{d}),
\end{align}
where $n_d=\sum_{k}\langle d_{k\sigma}^{\dag}d_{k\sigma}\rangle$. The functions $\mathfrak{F}_{i}(\Delta,\Delta_{nl},n_{d})$ and
$\mathfrak{G}_{i}(\Delta,\Delta_{nl},n_{d})$ are given by
\begin{align}\label{19}
\mathfrak{F}_{i}(\Delta,\Delta_{nl},n_{d})\!=\!\left[\frac{\omega_{ik}^2(\xi_{k}\!+\!U\frac{n_{d}}{2})\!-\!\xi_{k}U^2\left(1
\!-\!\frac{n_{d}}{2}\right)^2}{\omega_{ik}}\right.\nonumber\\
\left.-\frac{\Delta(U\Delta_{nl}-\mu\Delta)}{\omega_{ik}}\right]\tanh\left(\frac{\beta\omega_{ik}}{2}\right)
\end{align}
and
\begin{align}\label{20}
\mathfrak{G}_{i}(\Delta,\Delta_{nl},n_{d})=\frac{1}{\omega_{ik}}\left\{
\Delta\left[\omega_{i}^2+\xi_{k}U\left(1-\frac{n_{d}}{2}\right)\right]\right.\nonumber\\
\left.-U(U\Delta_{nl}-\mu\Delta)\left(1-\frac{n_d}{2}\right)\right\}\tanh\left(\frac{\beta\omega_{ik}}{2}\right).
\end{align}
Above $\beta=1/k_{B}T$ with $k_{B}$ being the Boltzmann
constant and $T$ the temperature.

As will be discussed in more detail below, not all values
of the parameters in the equations above give rise to real
values for the energies $\omega_{(1,2)k}$. Figure \ref{fig1} shows that only for $\mu/U < 1$ the energies are real for finite values of the
superconducting order parameter (we take $\Delta= 0$). For
a fixed $\mu/U < 1$ in Fig. \ref{fig1} as $\Delta_{nl}$ increases above the lines in the figure, the energies become complex. A possible
interpretation for energies with an imaginary part is the
appearance of a metastable state associated with a first
order phase transition and phase separation.

\section{Strong repulsive interactions}

\begin{figure}[th] \centering
\includegraphics[angle=0,scale=1.0,height=7.2cm]{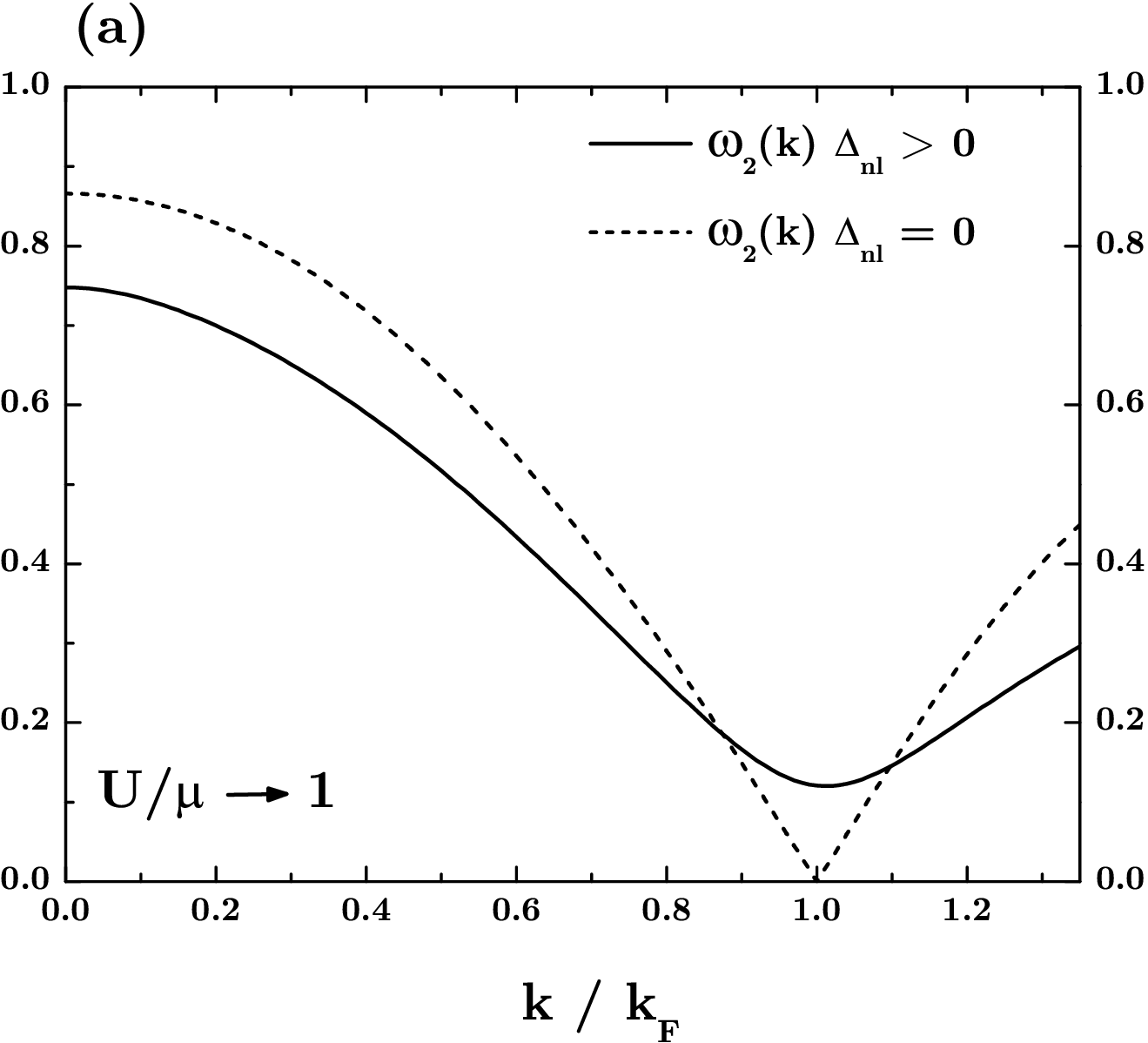}
\includegraphics[angle=0,scale=1.0,height=7.0cm]{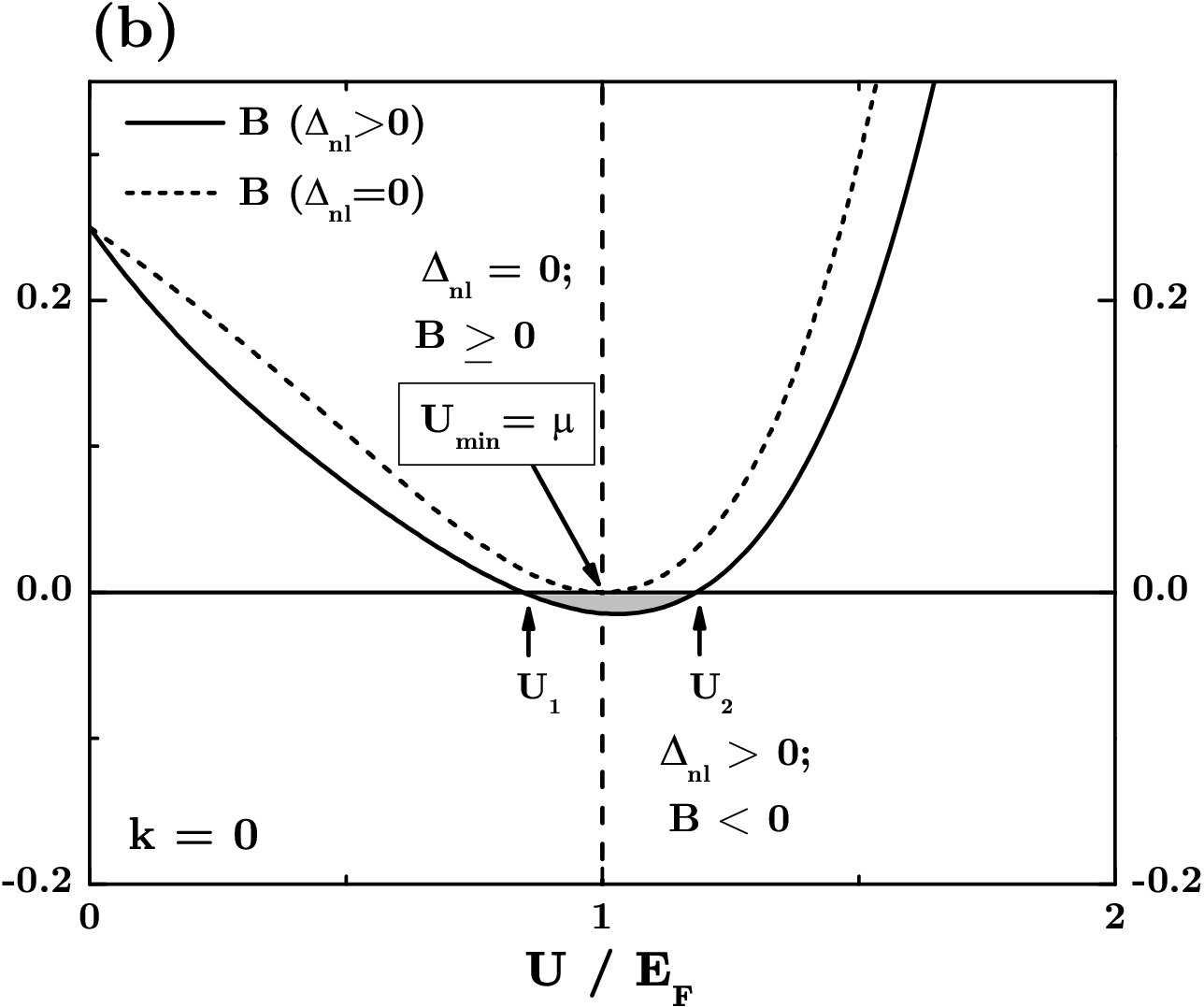}
\caption{(a) Dispersions relations of $\omega_{2}(k)$ of the \emph{quasi}-particles excitations for $U/\mu\rightarrow 1$ and $\Delta_{nl}>0$ and $\Delta_{nl}=0$. (b) It is shown for $k=0$ and finite $\Delta_{nl}$ the $U$ region with negative values of $B(k)$ that gives non-real $\omega_{1,2}(k)$ energies. For $\Delta_{nl}=0$ the condition in Eq. (\ref{21}) $U_{min}=\mu$ is shown. For all figures we used $\mu/E_{F}=1.0$ and $n_{d}=0.5$.} \label{fig2}
\end{figure}

In this system described by a conventional repulsive Hubbard model, we may expect that exists a minimum value of $U$ to give rise to a superconducting state. We consider that for finite and strong $U$ the local gap amplitude $\Delta \approx 0$ remaining only the non-local $\Delta_{nl}$ amplitude. At this point, it is necessary to analyze the spectra of \emph{quasi}-particle energies to check the range of the parameters, including $U$, which is consistent with a superconducting ground state. We plot in Fig.\ref{fig2}(a) the dispersion relations of the \emph{quasi}-particles and their coefficients $A(k)$ and $B(k)$  in Eq.(\ref{15}) to observe the role of the repulsive potential in the superconducting ground state ($\Delta_{nl} > 0$). The coefficient $A(k)$ in Eq.(\ref{15}) is always positive for all values of $U$. However $B(k)$ shows a negative region for some values of $U$ leading to non-real values of $\omega_{1,2}(k)$. These negative values of $B(k)$ appear for $U/E_F < 1$. For $U/E_F > 1$, the \emph{quasi}-particles energies $\omega_{1,2}(k)$ become continuous real functions for all range of $k$ space.

Indeed, to obtain $\omega_{ik}\in\Re$, we impose that $B(k)\geq0$ for
all $k$. Analyzing the case $\Delta_{nl} = 0$ appropriate to study
the system on the transition, we choose $k = 0$, such that,
$\xi_{k}(k=0)=-\mu$ to represent the most unfavorable case of
the strong interaction regime. Solving for the inequality
$B(k)\geq0$, we obtain

\begin{equation}\label{21}
U/\mu \geq 1,
\end{equation}

i.e., the minimum value for repulsive potential that gives a
continuous energy with physical meaning must be $U_{min}=U_c=\mu$.

To have a better comprehension about the $U_{min}$, we plot in the
Fig.\ref{fig2}(b) the \emph{quasi}-particles coefficient $B(k)$ as a function of $U$ for $k=0$. The analytical result of the inequality
for $B(k)\geq0$ and finite $\Delta_{nl}$ gives two solutions that
excludes the possibility to find non-real energies. However, in the
range between $0$ and $U_{1}$ is observed non-real energies for
$k>0$ as seen in Fig.\ref{fig2}(a) for $U/E_{F}=0.5$. For values from
$U_{2}$ to infinity all parameters are real. Then is reasonably
assume that for finite $\Delta_{nl}$, $U_{2}$ have a physical
meaning of minimum value of repulsive potential that stabilize a
superconducting state. Notice that the value of $U_2$
($\Delta_{nl}>0$) is bigger than $U_{min}$ obtained from
$\Delta_{nl}=0$. After these considerations, we analyze the self-consistent equations to characterize this unconventional superconducting order parameter driven by the electronic kinetic energy. Similar aspects of kinetically driven pairing in strongly correlated systems were investigated by Yanagisawa\cite{yanagisawa} in the context of the two-dimensional Hubbard model.

For $\Delta=0$ it remains a system of two equations and two
variables to be solved. Eqs. (\ref{16}) and (\ref{18}) becomes respectively
\begin{align}\label{22}
n_{d}\!=\!\frac{1}{2}\underset{k}\sum\left[1\!-\!\frac{1}{(\omega_{1k}^2\!-\!\omega_{2k}^2)}
\underset{i=1}{\overset{2}{\sum}}(-1)^{i-1}\frak{F}_{i}(\Delta_{nl},n_{d})\right]
\end{align}
and
\begin{align}\label{23}
1=U^2\left(1\!-\!\frac{n_{d}}{2}\right)\underset{k}\sum\frac{\epsilon_{k}}{2(\omega_{1k}^2\!-\!\omega_{2k}^2)}
\underset{i=1}{\overset{2}{\sum}}(\!-1)^{\!i-1}
\frak{G}_{i}(\Delta_{nl},n_{d}),
\end{align}
where the functions $\frak{F}_{i}(\Delta_{nl},n_{d})$ and
$\frak{G}_{i}(\Delta_{nl},n_{d})$ now are given by
\begin{align}\label{24}
\frak{F}_{i}(\Delta_{nl},n_{d})\!=\!\frac{\omega_{ik}^2(\xi_{k}\!+\!U\frac{n_{d}}{2})\!-\!\xi_{k}U^2(1
\!-\!\frac{n_{d}}{2})^2}{\omega_{ik}}\tanh\left(\frac{\beta\omega_{ik}}{2}\right)
\end{align}
and
\begin{align}\label{25}
\frak{G}_{i}(\Delta_{nl},n_{d})=\frac{1}{\omega_{ik}}\tanh\left(\frac{\beta\omega_{ik}}{2}\right).
\end{align}

\begin{figure}[th] \centering
\includegraphics[angle=0,scale=1.0,height=7.0cm]{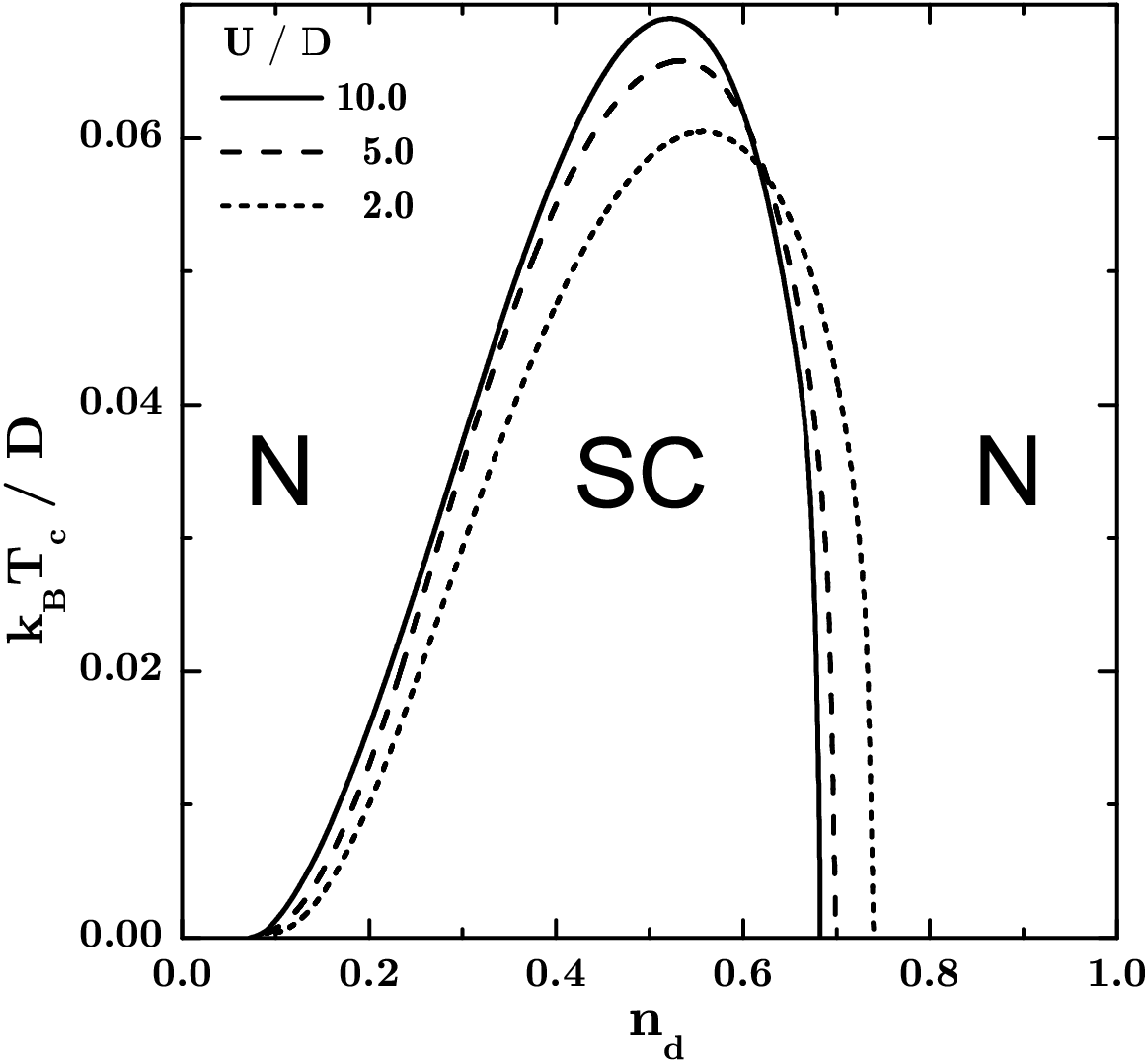}
\caption{The $T_{c}\times n_d$ phase diagram for an electron gas in two dimensions. For strong repulsive interactions the highest value of $T_c$ is observed around $n_d=0.5$}\label{fig3}
\end{figure}

Since we are interested in the 2D system to solve the self-consistent equations we introduce a constant density of states with a fixed width $D$ that reproduces the 2D scenario, i.e., we take, $\rho(\omega)=\sum_{k}\delta(\omega-\epsilon_k) = 1/D$ for $0 < \omega < D$ and $\rho(\omega)=0$ for $\omega > D$. Then, equations (\ref{22}) and (\ref{23}) can be rewritten as,

\begin{align}\label{26}
n_{d}\!=\!\frac{1}{2}\int_{-\overline{\mu}}^{1-\overline{\mu}}dx\left[1\!-\!\frac{1}{(\overline{\omega}_{1x}^2\!-\!\overline{\omega}_{2x}^2)}
\underset{i=1}{\overset{2}{\sum}}(-1)^{i-1}\frak{F}_{i}(\overline{\Delta}_{nl},n_{d})\right]
\end{align}
and
\begin{align}\label{27}
1=\gamma(\overline{U},n_d)\int_{-\overline{\mu}}^{1-\overline{\mu}}\frac{(x+\overline{\mu})dx}{(\overline{\omega}_{1x}^2\!-\!\overline{\omega}_{2x}^2)}
\underset{i=1}{\overset{2}{\sum}}(\!-1)^{\!i-1}
\frak{G}_{i}(\overline{\Delta}_{nl},n_{d}),
\end{align}
where $\gamma(\overline{U},n_d)=(1\!-\!n_{d}/2)\overline{U}^2/2$.
The variable $x\equiv(\omega-\mu)/D$, and the
overline in the parameters means that they were normalized
by the band width $D$. Is important to highlight that in the Eqs.
(\ref{26}) and (\ref{27}) the functions
$\frak{F}_{i}(\overline{\Delta}_{nl},n_{d})$ and
$\frak{G}_{i}(\overline{\Delta}_{nl},n_{d})$ were normalized by $D$
and now are functions of $x$ and by consequence, functions of
$\overline{\omega}_{1x}$ and $\overline{\omega}_{2x}$. These can be
easily obtained performing a normalization of Eq. (\ref{15}). Thus,
solving self-consistently Eqs. (\ref{26}) and (\ref{27}) we can obtain the phase diagram of the system. Fig.
\ref{fig3} shows the $T_c\times n_d$ phase diagram for several
values of $U$. For strong repulsive interactions ($\overline{U}>10$)
the highest value of $T_c$ is observed around $n_d=0.5$.

\begin{figure}[th] \centering
\includegraphics[angle=0,scale=1.0,height=7cm]{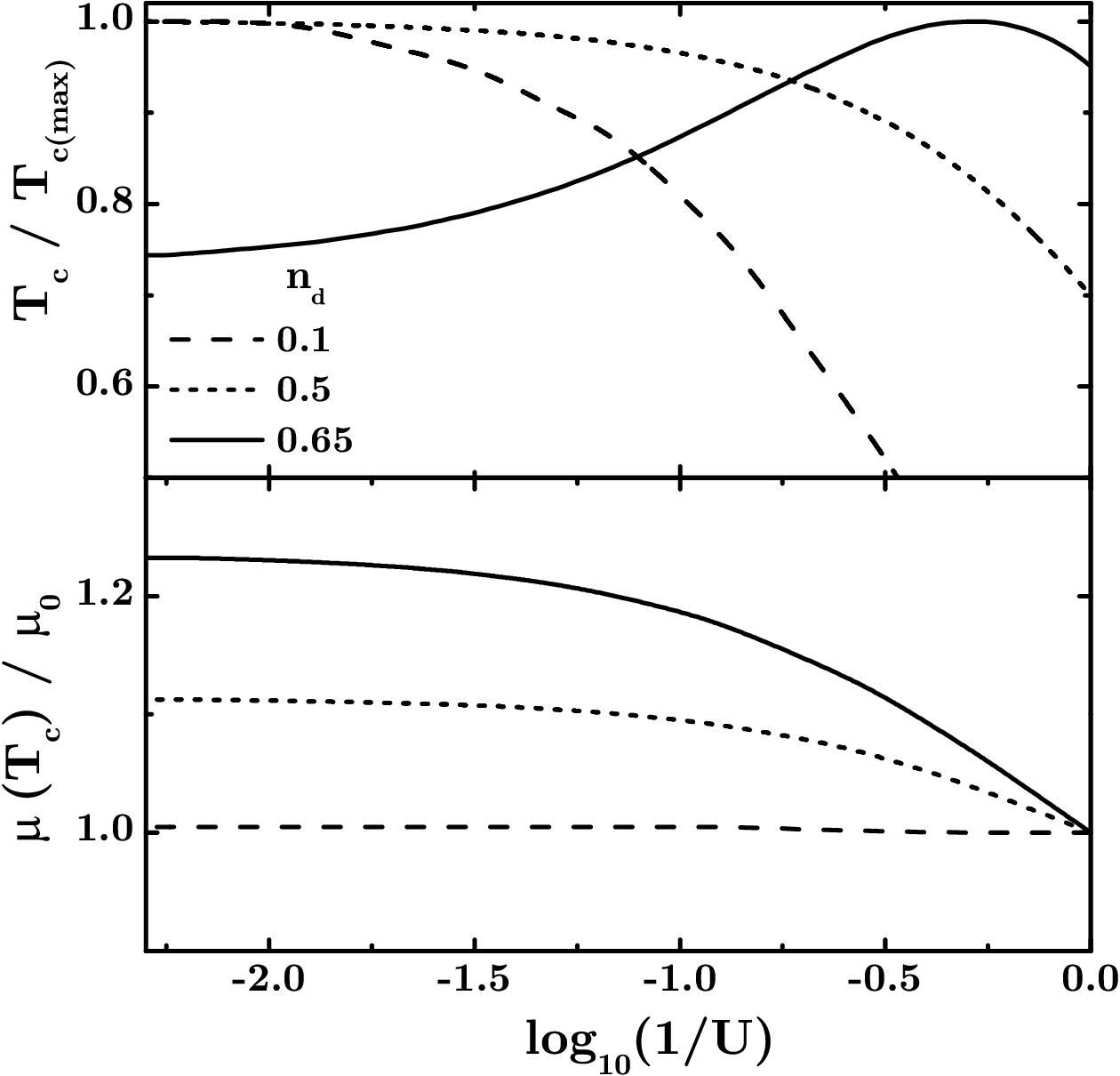}
\caption{$T_c$ and chemical potential ${\mu}(T_c)$ as function of $U$ in logarithmic scale for strong repulsive interaction. Notice that $U$ is normalized by the band width $D$ and $\mu_0=\mu(T_c,U/D=1)$.}\label{fig4}
\end{figure}

Notice also that, independent of the value of $U$, even larger than $U_c$, there is no superconducting solution for $0 < n_d < 0.08$ implying that it is necessary a minimum concentration of electrons to reach the SC state. Moreover, qualitatively the phase diagrams practically does not change as a function of $U$ and always shows a dome shape.

Fig. \ref{fig4} shows the critical lines separating normal and
superconducting phases for the case of strong repulsive
interactions. For $0.1 < n_d < 0.5$ we observe an identical
behavior: $T_c$ monotonically grows with a logarithmical
dependence on $U^{-1}$, and starts to become constant when
$U \sim 10$ . In this range of $n_d$ the maximum value of the
critical temperature $T_{c(max)}$ coincides with the $T_c$ value
for $U\rightarrow\infty$. However, for $n_d > 0.5$ the $T_{c(max)}$ no more coincides with $T_c$ value for $U\rightarrow\infty$. In this scenario $n_d$ and $U$ start to act in detriment of superconductivity. Nevertheless, for all cases, $T_c$ saturates in the strong $U$ limit. This feature is quite similar to that observed in the BEC scenario for strong attractive interactions. However, here the chemical potential potential does not decrease, vanishes and becomes negative with increasing interaction as in the BEC case. On the contrary, it increases or becomes constant with increasing $U$. Notice that this is not unexpected since, with repulsive interactions there is no formation of bosonic pairs that could condense.

\begin{figure}[th] \centering
\includegraphics[angle=0,scale=1.0,height=9.0cm]{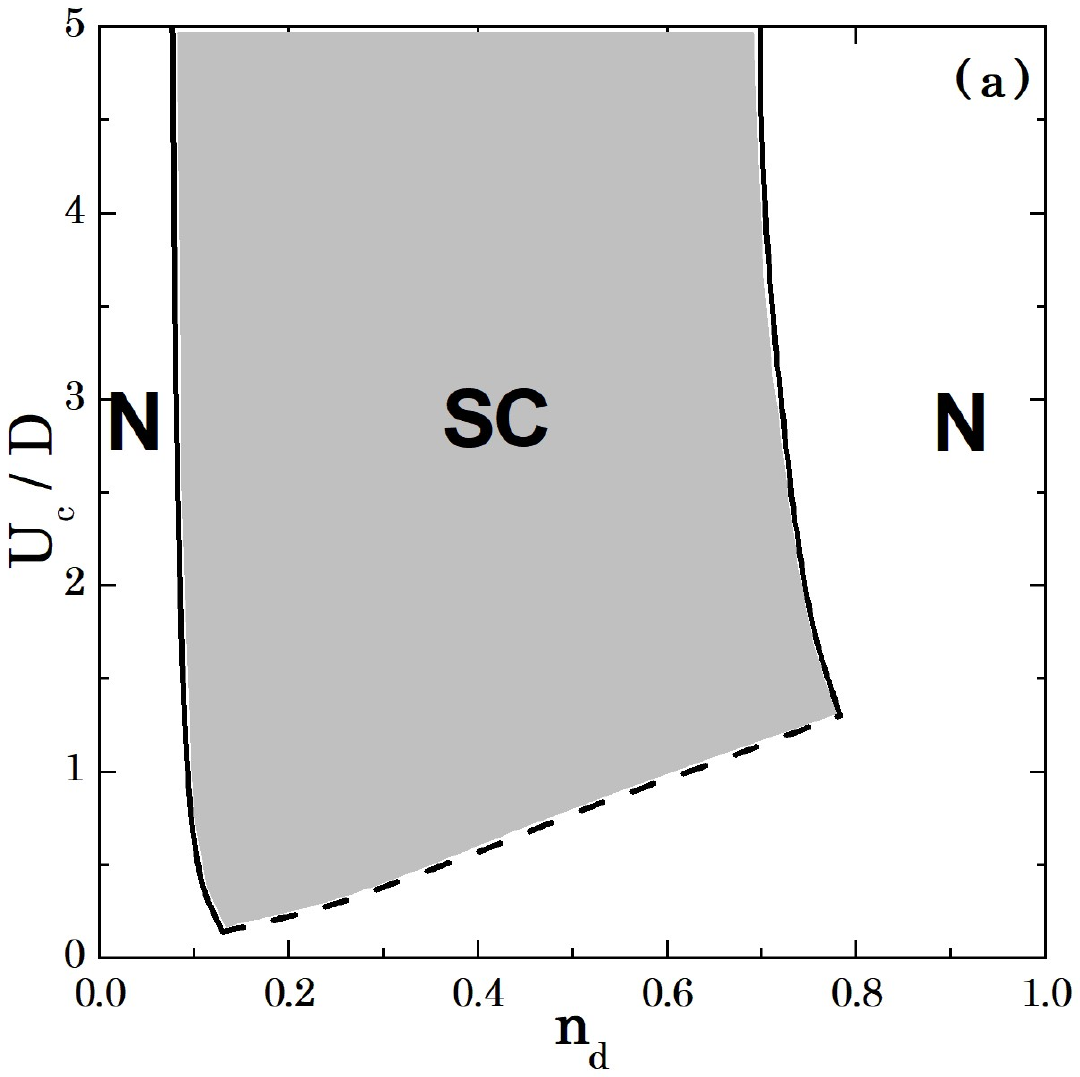}
\includegraphics[angle=0,scale=1.0,height=8.5cm]{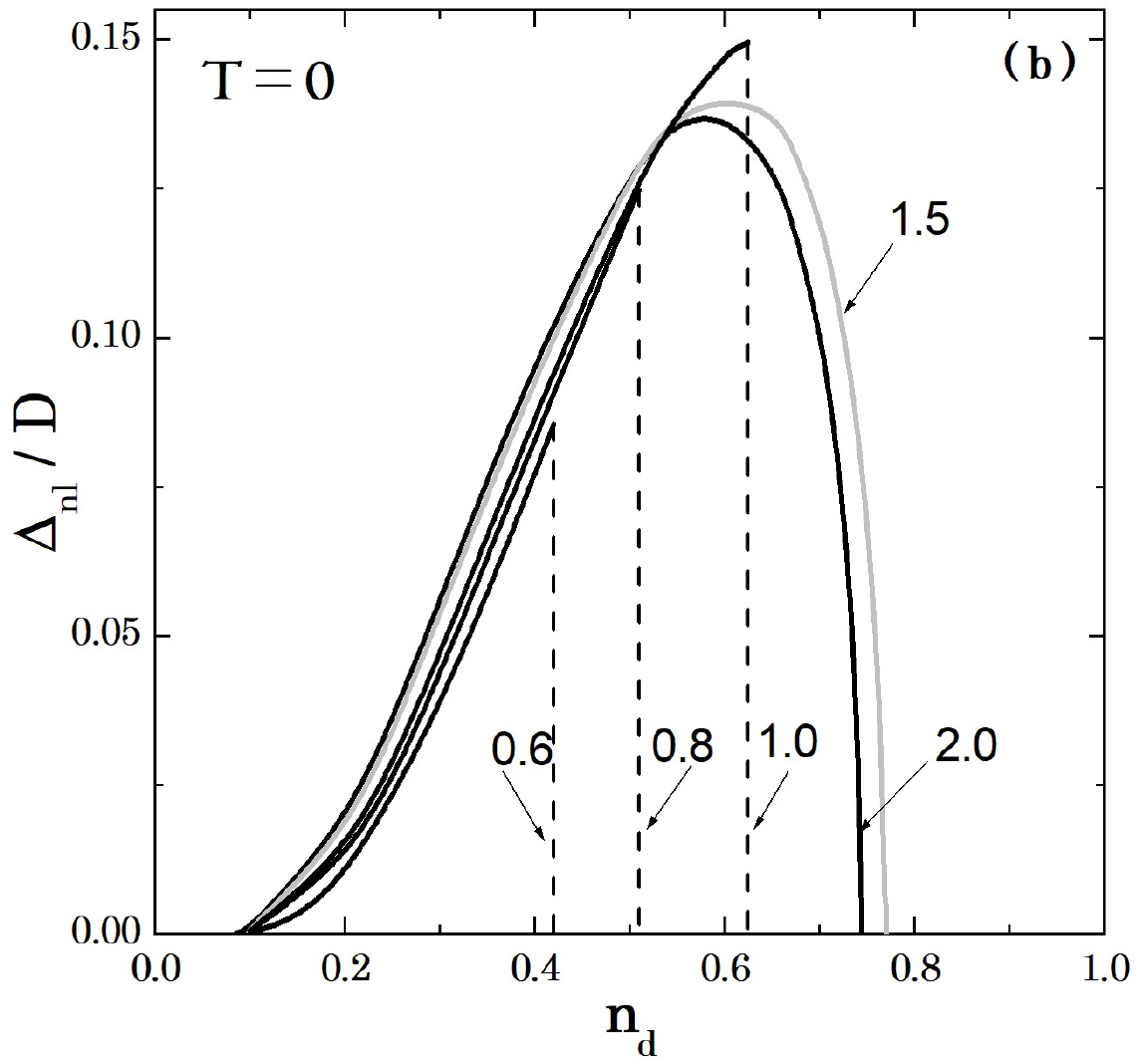}
\caption{a) The ground state ($T=0$) phase diagram $U_c \times n_d$ for an electron gas in 2D with a constant density of states $\rho(\omega)=1/D$. The dashed line represents the first-order phase transition line. b) The non-local order parameter $\Delta_{nl}$ as a function of $n_d$ for several values of $U/D$. In this case, we observe both second-order and first-order phase transitions.}\label{fig5}
\end{figure}

Finally, in the Fig. \ref{fig5}(a) we show the ground state ($T=0$)
phase diagram of the system. Once again is possible to observe the
saturation behavior when the critical lines becomes parallel and
constant for $U\rightarrow\infty$. The bottom dashed line represents the first-order phase transition line obtained by numerical analysis of non-local order parameter $\Delta_{nl}$ in function of $n_d$ for several values of $U/D$ as shown in Fig. \ref{fig5}(b). It is important to emphasise that these results were obtained for a constant density of states $\rho(\omega)=1/D$. We initially worked in the strong repulsion $U$ limit, where the local gap $\Delta$ can be neglected. We then extended the analysis to the regime ($U \sim D$). For $U/D$ = 2.0 and 1.5, Fig. \ref{fig5}(b) shows a continuous SC–N transition from the behavior of $\Delta_{nl}$, characteristic of a second-order phase transition. For $U/D$ = 1.0, 0.8, and 0.6, however, $\Delta_{nl}$ vanishes discontinuously at a critical $n_d$, indicating a first-order transition. These points form the dashed critical line at the bottom of the $U_c \times n_d$ phase diagram in Fig. \ref{fig5}(a). This occurs as a consequence of Eq. (\ref{21}): in the weak $U$ regime, for some $n_d$ values, the condition for real physical solutions breaks down because $U < \mu$.

\section{Discussions about strong attractive interactions}

In the previous sections we discussed about the superconductivity
that arises from repulsive interactions. We have observed a $T_c$ saturation for strong repulsive $U$, but in this case this is not related to the formation of effective bosons from Cooper pairs. In this case we do not observe a chemical potential shift from finite values to zero that indicates the disappearance of fermions and gives rise to a creation of effective bosons and their condensation. In this section we propose to discuss the attractive Hubbard
model, but now using the Hubbard-I decoupling method.
We expect to find more satisfactory results in comparison
with the ordinary Hartree-Fock mean-field technique, at
least for strong interactions. The conventional mean-field
or Hartree-Foch approximation provides good results for
the SC ground state. However, without including Gaussian
fluctuations in the number equation it is not enough
to describe the crossover BCS-BEC for the critical temperature
$T_c$.

In our treatment for strong attractive interactions we consider now only the local gap equation ignoring the non-local term i.e. $\Delta_{nl} \approx 0$. This is a reasonably approach for example for SC materials that show a very short coherence length, where the pairs are tightly bound.

\begin{figure}[th] \centering
\includegraphics[angle=0,scale=1.0,height=7.0cm]{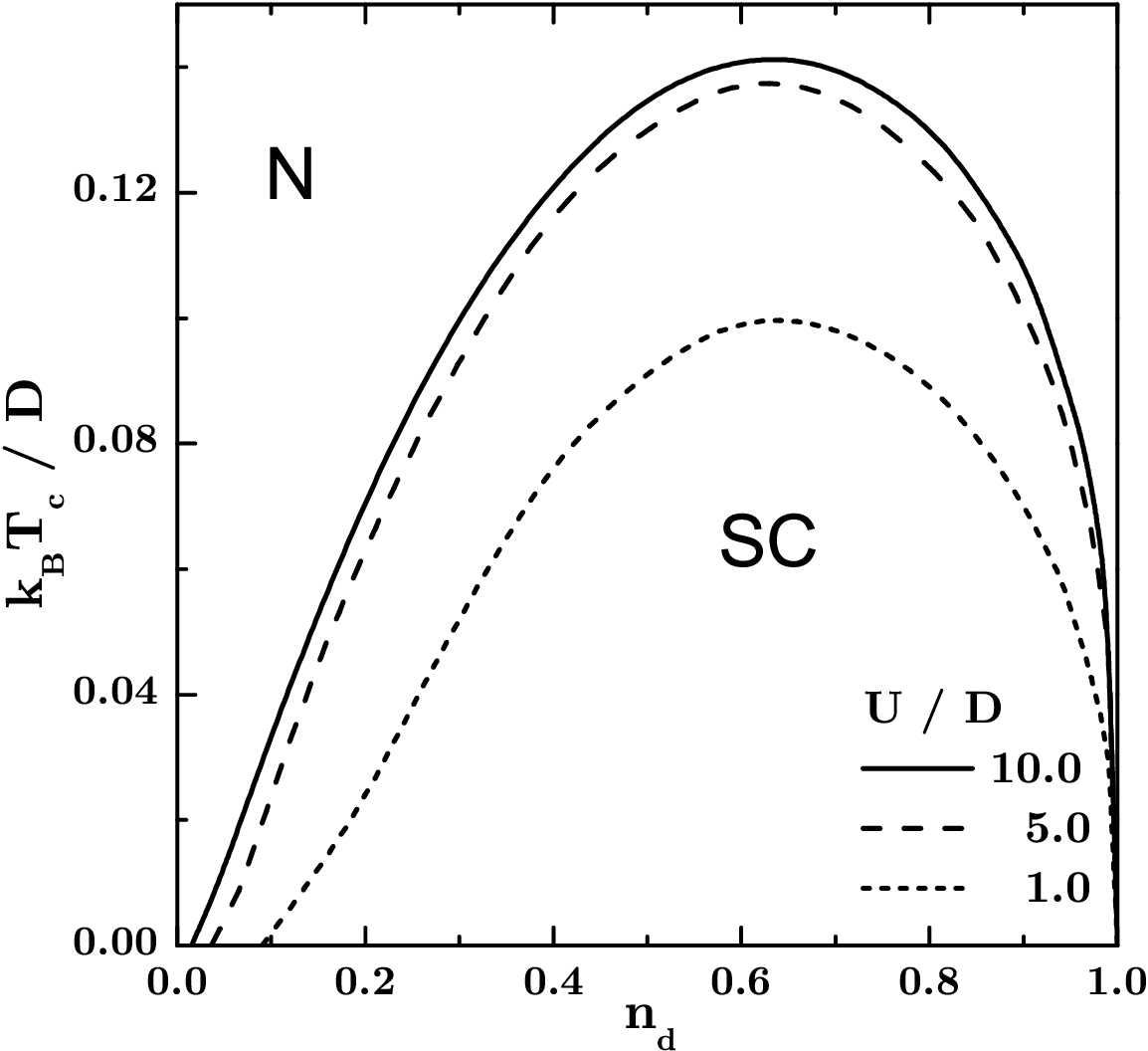}
\caption{The $T_c \times n_d$ phase diagram for attractive
interactions. For this case the SC state is only destroyed when
$n_d\rightarrow 1$.}\label{fig6}
\end{figure}

Thus for local pairing the gap equation is given by Eq. (\ref{17}) where now we must take into account the fact that $U<0$. We get,
\begin{align}\label{28}
1=\frac{U}{2}\underset{k}\sum\frac{1}{(\omega_{1k}^2\!-\!\omega_{2k}^2)}
\underset{i=1}{\overset{2}{\sum}}(\!-1)^{\!i-1}
\frak{G}_{i}(\Delta,n_{d}),
\end{align}
where
\begin{align}\label{29}
\frak{G}_{i}(\Delta,n_{d})=\frac{\omega_{ik}^2-U\epsilon_{k}(1-n_d/2)}{\omega_{ik}}\tanh\left(\frac{\beta\omega_{ik}}{2}\right).
\end{align}

For the number equation we have a similar result obtained in Eq.
(\ref{22}), but with the characteristic function,

\begin{align}
&\frak{F}_{i}(\Delta,n_{d})=\nonumber\\
&\!\frac{\omega_{ik}^2(\xi_{k}\!-\!Un_{d}/2)\!-\!\xi_{k}U^2(1
\!-\!n_{d}/2)^2\!+\!\mu\Delta^2}{\omega_{ik}}\tanh\left(\frac{\beta\omega_{ik}}{2}\right).
\end{align}

Following the same procedure of the former case we rewrite both gap
and number equations using a constant density of states. In
Fig. \ref{fig6} we observe the $T_c \times n_d$ phase diagram. In
this case we observe that for any strength of the interaction the SC phase is destroyed only for $n_d=1$. In this case the motion of Cooper pairs is avoided acting effectively in detriment of superconductivity for several values of $U/D$\cite{sacra}.

\begin{figure}[th] \centering
\includegraphics[angle=0,scale=1.0,height=7.0cm]{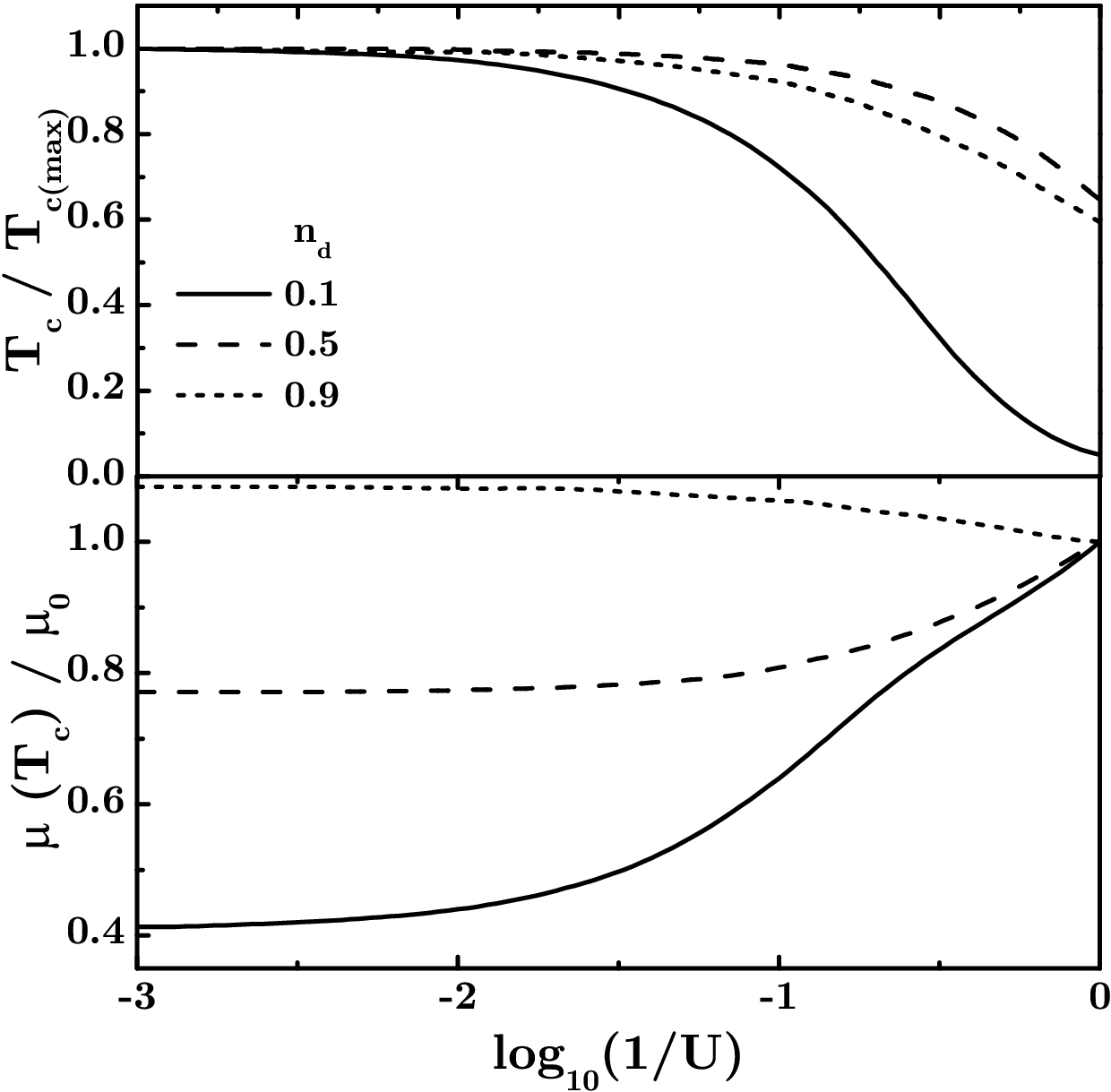}
\caption{$T_c$ and chemical potential ${\mu}(T_c)$ as functions of attractive $U$. $U$ is normalized by the
band width $D$ and $\mu_0=\mu(T_c,U/D=1)$}\label{fig7}
\end{figure}

In Fig. \ref{fig7} we show the critical lines for strong attractive interactions. The $T_c$ curves show an identical behavior for all values of occupation: $T_c$ monotonically grows with a logarithmical dependence on $U^{-1}$, and starts to become constant when $U \sim 10$. In this case, for all values of $n_d$, the maximum value of the critical temperature $T_{c(max)}$ coincides with $T_c$ for $U \rightarrow \infty$ where a saturation is observed. To verify if this saturation indicates a BEC of Cooper pairs, it is necessary to analyze the chemical potential $\mu(T_c)$. In the lower graph of Fig. \ref{fig7} we plot the $\mu(T_c)/\mu_0$ $\times$ $log_{10}(1/U)$ curves. We define $\mu_0 = \mu(T_c,U/D = 1)$ to normalize $\mu(T_c)$. For $n_d = 0.1$, $\mu(T_c)$ decreases whereas $U$ grows. This is a typical behavior observed when the Cooper pairs condensate\cite{samelo,dinola,dinola2,rand}. However, even for $U\rightarrow\infty$, $\mu(T_c)$ does not vanish or becomes negative. This fact seems to indicate that even in the strong coupling limit not all Cooper pairs condensate, i.e. in the SC ground state can be found ordinary Cooper pairs and the effective bosons composed by tightly bound electrons. Notice that for $n_d = 0.9$, $\mu(T_c)$ increases. Thus for $n_d \rightarrow 1$ the SC state mediated by attractive interactions behaves similarly to the repulsive case.

For a general comparison we had performed the same procedure analyzing $B(k)$. Now we obtain another condition
to exclude non-real values of energies. The inequality $B(k=0)\geq 0$ lead us to $U/\mu\leq-1$, i.e. the minimum
value for attractive interactions in the strong coupling limit must be $U_{min}=-\mu$. This fact indicates that also
for $U < 0$ there exists the same effect of phase separation, associated with a first order transition as found in
the repulsive case. This is associated with a soft mode appearing at the Fermi surface as in the previous case of
$U > 0$.

\section{Conclusions}

In this paper we studied superconductivity in a repulsive Hubbard model. We restrict our analysis for the 2D electron gas scenario. The Hubbard-I method provides a non-local gap amplitude even for a local repulsion besides a local gap amplitude. The symmetry of the non-local gap is extended s-wave with a similar k-dependence of the dispersion of the bare electrons in the gas. For strong repulsive interactions the local amplitude vanishes and the non-local term plays the role of SC order parameter. In this scenario, we obtain the minimum value of repulsion $U_{min}$ that gives rise to a stable SC state. For finite temperatures, we observe a saturation of $T_c$ for all occupations $n_d$ around $U/D \sim 10$. In this repulsive case the chemical potential increases as the interaction $U$ grows. This is an evidence that the system keeps its fermionic character without formation of effective bosons that for attractive interactions gives rise to a condensate.

We treated also the attractive case to compare with the repulsive one. In this case we only focused in the strong
coupling limit, and by consequence we neglected the nonlocal terms and considered only the local gap amplitude.
For finite temperatures the SC state is only destroyed for $n_d = 1$ for several values of attractive $U$. This fact
differs of the repulsive case where it was observed a critical $n_d \sim 0.6$ for strong repulsive interactions. However,
the attractive case shows the decrease of the chemical potential for $n_d = 0.1$ and $n_d = 0.5$, and as expected
the system now are composed by effective bosons. Nevertheless, the chemical potential does not vanish for the
representative values of $n_d = 0.1$ and $n_d = 0.5$ saturating in a finite value for $U \rightarrow \infty$. An interesting point is that for $n_d = 0.1$ the chemical potential saturates in a value much lower than for $n_d = 0.1$ case. We presume that for
finite concentrations but lower than $n_d = 0.1$ all fermions condensate. In the case of doping, with $\delta = 1 - n_d$ \cite{Sarasua} this means that all Cooper pairs can condense in the high doping limit, i.e. $\delta \sim 1$.

Finally, we propose that there is a phenomenon of phase separation in these systems. This occurs for $\mu/U \geq
1$ and is associated with the appearance of an imaginary part in the energy of the quasi-particles. This signals the
unstable character of the superconductivity phase in the region of the phase diagram where this imaginary part is
finite.

\acknowledgments Minos A. Neto thank the Brazilian Agency CNPq for partial financial support.


\begin{thebibliography}{99}

\bibitem{bcs}J. Bardeen, L. N. Cooper and J. R. Schrieffer, \textit{Phys. Rev.} \textbf{106},
162 (1957).

\bibitem{HTSC}J. G. Bednorz and K. A. M\"{u}ller, \textit{Z. Phys. B} \textbf{64}, 189 (1986)

\bibitem{Gough}C. E. Gough \textit{et al.}, \textit{Nature} \textbf{326}, 855 (1987).

\bibitem{Anderson}P. W. Anderson, \textit{Science} \textbf{235}, 1196 (1987).

\bibitem{nozieres}P. Nozi\`{e}res and S. Schmitt-Rink, \textit{J. Low Temp.
Phys.} \textbf{59}, 195 (1985).

\bibitem{samelo}C. A. R. S\'{a} de Melo, M. Randeira and J.R.
Engelbrecht, \textit{Phys. Rev. Lett.} \textbf{71}, 3202 (1993).

\bibitem{Beenen}J. Beenen and D. M. Edwards, \textit{Phys. Rev. B} \textbf{52}, 13636
(1995).

\bibitem{Sarasua}L. G. Sarasua, \textit{Phys. Scr.} \textbf{84}, 045706
(2011).

\bibitem{Lisandrini}F. T. Lisandrini \textit{et al.}  arXiv:2510.09363v1 [cond-mat.str-el] (2025)

\bibitem{bastide}C. Bastide, C. Lacroix and A. da Rosa Sim\~{o}es,
\textit{Physica C} \textbf{159}, 347 (1989).

\bibitem{dinola}F. Din\'{ola} Neto, M. A. Continentino and C.
Lacroix, \textit{Physica C} \textbf{485}, 75 (2013).

\bibitem{caixeiro}E. S. Caixeiro and A. Troper, \textit{Phys. Rev. B} \textbf{82}, 014502
(2010).

\bibitem{hubbard}J. Hubbard, \textit{Proc. Roy. Soc. A} \textbf{276}, 238
(1963).

\bibitem{japiassu}G. M. Japiassu, M. A. Continentino, and A. Troper, \textit{Phys. Rev. B} \textbf{45}, 2986 (1992).

\bibitem{yanagisawa}T. Yanagisawa, \textit{Phys. Lett A} \textbf{403}, 127382 (2021).

\bibitem{sacra}P. D. Sacramento, J. Apar\'{i}cio and G. S.
Nunes, \textit{J. Phys.: Cond. Matt.} \textbf{22} 065702 (2010).

\bibitem{dinola2}F. Din\'{o}la Neto, M. A. Continentino and C.
Lacroix, \textit{J. Phys.: Condens. Matter} \textbf{22}, 075701 (2010).

\bibitem{rand}M. Randeria, J. Duan, and L. Shieh, \textit{Phys.
Rev. B} \textbf{41}, 327 (1990).

\end{thebibliography}
\end{document}